\documentclass[aps,prl,reprint,twocolumn,superscriptaddress,floatfix,nofootinbib,longbibliography]{revtex4-2}
\usepackage{graphicx,amsmath,amsfonts,amssymb,amsthm,xr}
\usepackage{epsfig,amsmath,amssymb,color,dsfont,upgreek,physics}
\usepackage{mathrsfs}
\usepackage{mathtools}
\usepackage{bbold}
\usepackage{float}
\usepackage[caption=false]{subfig}

\usepackage[bookmarks=true,colorlinks,linkcolor=OrangeRed,urlcolor=NavyBlue,citecolor=RoyalBlue]{hyperref}
\usepackage[dvipsnames]{xcolor}
\usepackage{orcidlink}

\usepackage{graphicx}
\usepackage{dcolumn}
\usepackage{bm}
\usepackage{color}

\begin{document}

\preprint{}
\title{Bosonic vs.~Fermionic Matter in Quantum Simulations of $2+1$D Gauge Theories}

\author{N.~S.~Srivatsa}
\affiliation{Max Planck Institute of Quantum Optics, 85748 Garching, Germany}
\affiliation{Munich Center for Quantum Science and Technology (MCQST), 80799 Munich, Germany}

\author{Jesse J.~Osborne${}^{\orcidlink{0000-0003-0415-0690}}$}
\affiliation{School of Mathematics and Physics, The University of Queensland, St.~Lucia, QLD 4072, Australia}

\author{Debasish Banerjee${}^{\orcidlink{0000-0003-0244-4337}}$}
\affiliation{School of Physics and Astronomy, University of Southampton, University Road, SO17 1BJ, UK}

\author{Jad C.~Halimeh${}^{\orcidlink{0000-0002-0659-7990}}$}
\email{jad.halimeh@physik.lmu.de}
\affiliation{Max Planck Institute of Quantum Optics, 85748 Garching, Germany}
\affiliation{Department of Physics and Arnold Sommerfeld Center for Theoretical Physics (ASC), Ludwig Maximilian University of Munich, 80333 Munich, Germany}
\affiliation{Munich Center for Quantum Science and Technology (MCQST), 80799 Munich, Germany}

\date{\today}

\begin{abstract}

Quantum link models extend lattice gauge theories beyond the traditional Wilson formulation and present promising candidates for both digital and analog quantum simulations. Fermionic matter coupled to $\mathrm{U}(1)$ quantum link gauge fields
has been extensively studied, revealing a phase diagram that includes transitions from the columnar phase in the quantum dimer model to the resonating valence bond phase in the quantum link model, potentially passing through a disordered liquid-like phase. In this study, we investigate the model coupled to hardcore bosons and identify a similar phase structure, though with a more intricate mixture of phases around the transition. Our analysis reveals that near the transition region, a narrow and distinct ordered phase emerges, characterized by gauge fields forming plaquette configurations with alternating orientations, which is then followed by a thinner, liquid-like regime. This complexity primarily stems from the differences in particle statistics, which manifest prominently when the matter degrees of freedom become dynamic. Notably, our findings suggest that bosons can effectively replace fermions in lattice gauge theory simulations, offering solutions to the challenges posed by fermions in both digital and analog quantum simulations.

\end{abstract}

\maketitle

\textbf{\emph{Introduction.---}}
Lattice gauge theories (LGT) serve as a powerful framework for studying quantum field theories, particularly in the context of non-Abelian gauge theories such as those arising in quantum chromodynamics (QCD) \cite{Berges_review,QCD_review,Weinberg_book}. The conventional formulation of lattice gauge theories involves discretizing spacetime into a grid, where gauge fields are represented by link variables, typically chosen as matrices in a Lie group \cite{Rothe_book}.

\begin{figure}
    \includegraphics[width=6.2cm]{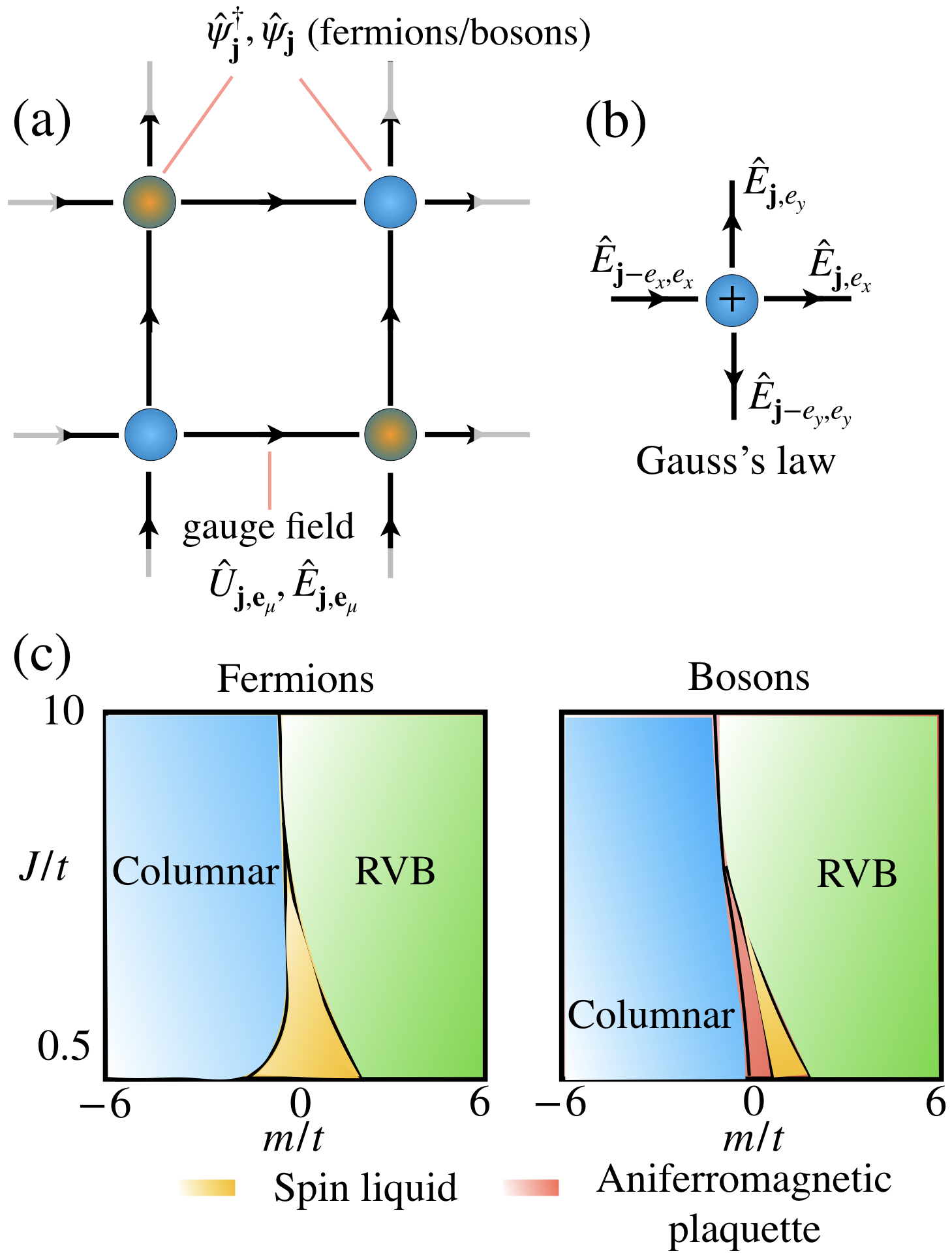}
    \caption{(a) Lattice structure for the $2+1$D $\mathrm{U}(1)$ quantum link model~\eqref{eq:QLM}. Matter fields (fermions/bosons) live on the sites and the gauge fields live on the links. (b) Illustration of Gauss's law condition Eq.~\eqref{gauss} for a positron on site $\mathbf{j}$. (c) Phase diagrams for fermions and hardcore bosons.}
    \label{illus}
\end{figure}

The conventional formulation has been extremely successful for the studies of the static properties of QCD, the nature of the associated quasiparticles, their decay, and scattering properties \cite{Fodor2012, Ding:2020rtq, DeGrand:2006zz, USQCD:2022mmc, Knechtli:2017sna}. However, it has been less successful in the study of dynamics and physics at finite densities because of the nature of the classical computations involved. Quantum computing \cite{NielsenChuang_book,Alexeev_review} and quantum simulation \cite{Georgescu_review} are an emerging venue that has the potential to fundamentally contribute in these directions but works better for microscopic theories which require small Hilbert spaces.

Quantum link models (QLMs) \cite{Horn:1981kk, Orland:1989st, Chandrasekharan1997,Wiese_review} offer a promising alternative to the standard formulation of lattice gauge theories, where the gauge invariance can be exactly realized with a local finite-dimensional Hilbert space, and are therefore ideal for implementations in quantum computing and simulation platforms \cite{Dalmonte_review, Pasquans_review, Zohar_review, aidelsburger2021cold, Zohar_NewReview, klco2021standard, Bauer_review, dimeglio2023quantum, Cheng_review, Halimeh_review, Cohen:2021imf, Lee:2024jnt, Turro:2024pxu}.
In QLMs, the gauge fields are represented by finite-dimensional quantum-mechanical operators (such as quantum spins) that reside on the links of the lattice. The size of the local Hilbert space of these operators can be tuned to asymptotically achieve the quantum field theory limit \cite{Buyens2017,Zache2021achieving,Halimeh2021achieving}. QLMs have attracted significant interest as they potentially offer new insights about novel phases of matter, phase transitions, and novel phenomena at excited states  \cite{Banerjee2013deconfined, Banerjee:2014wpa, Banerjee:2015pnt, Banerjee:2017tjn, Banerjee:2021tfo, Banerjee:2017tjn, Rico:2018pas,Brenes2018,Surace2020,Papaefstathiou2020, Sau2022, Banerjee2021, biswas2022scars,Hashizume2022,vandamme2022dqpt,Desaules2022weak,Desaules2022prominent,Halimeh2022robust, Desaules2024ergodicitybreaking,osborne2023disorderfreelocalization21dlattice,Sau:2023clm,osborne2024quantummanybodyscarring21d, Sau:2024uur,su2024particlecollider, tan2025,Jeyaretnam2025Hilbert}.

Fermionic matter interacting with dynamical gauge fields has been studied more extensively than bosonic matter, largely because the matter sector in the Standard Model of particle physics is inherently fermionic \cite{Weinberg_book}. The phase diagram of the $\mathrm{U}(1)$ QLM 
with fermionic matter has been explored, revealing a transition from a columnar phase, as in the quantum dimer model to a resonating valence bond (RVB) phase, as in the quantum link model as the mass of the matter fields is varied. Additionally, a disordered, liquid-like phase has been speculated near the transition between these phases \cite{Hashizume2022}. In this Letter, we investigate the phase diagram for the case of hardcore bosonic matter coupled to spin-$1/2$ quantum links, and compare it to the case with fermionic matter. We particularly focus on how differences due to particle statistics may affect physical observables in the ground state hoping to deepen our understanding of how the interplay between gauge fields and particle statistics influence the phase diagram. We observe that the phase structure of bosons is largely similar to that of fermions, with one important exception: in the region of small bare mass, where matter field hopping is enhanced, bosons exhibit a more intricate phase structure. 
This includes a probable first-order transition from a columnar phase to one where the gauge fields order, leading to the formation of plaquettes with opposing orientations. Following this, there may be a narrow spin-liquid phase, ultimately transitioning to an RVB phase. This is illustrated in Fig.~\ref{illus}. Our numerical analysis suggests that, deep in the confined phase, bosons are promising candidates for studying LGT phenomena such as confinement and string breaking. Given the impressive progress in LGT quantum simulation experiments \cite{Martinez2016,Klco2018,Goerg2019,Schweizer2019,Mil2020,Yang2020,Wang2021,Zhou2022,Wang2023,Zhang2023,Su2022,Ciavarella2024quantum,Ciavarella:2024lsp,de2024observationstringbreakingdynamicsquantum,liu2024stringbreakingmechanismlattice,Farrell:2023fgd,Farrell:2024fit,zhu2024probingfalsevacuumdecay,Ciavarella:2021nmj,Ciavarella:2023mfc,Ciavarella:2021lel,Gustafson:2023kvd,Gustafson:2024kym,Lamm:2024jnl,Farrell:2022wyt,Farrell:2022vyh,Li:2024lrl,Zemlevskiy:2024vxt,Lewis:2019wfx,Atas:2021ext,ARahman:2022tkr,Atas:2022dqm,Mendicelli:2022ntz,Kavaki:2024ijd,Than:2024zaj,Angelides2025first,alexandrou2025realizingstringbreakingdynamics}, including recently in two spatial dimensions \cite{cochran2024visualizingdynamicschargesstrings,gyawali2024observationdisorderfreelocalizationefficient,gonzalezcuadra2024observationstringbreaking2,crippa2024analysisconfinementstring2}, our findings are very timely and relevant in that they address a major question in the field: When is fermionic matter, with its usually significantly more expensive overhead, necessary when quantum simulating an LGT?

\begin{figure}
    \includegraphics[width=\linewidth]{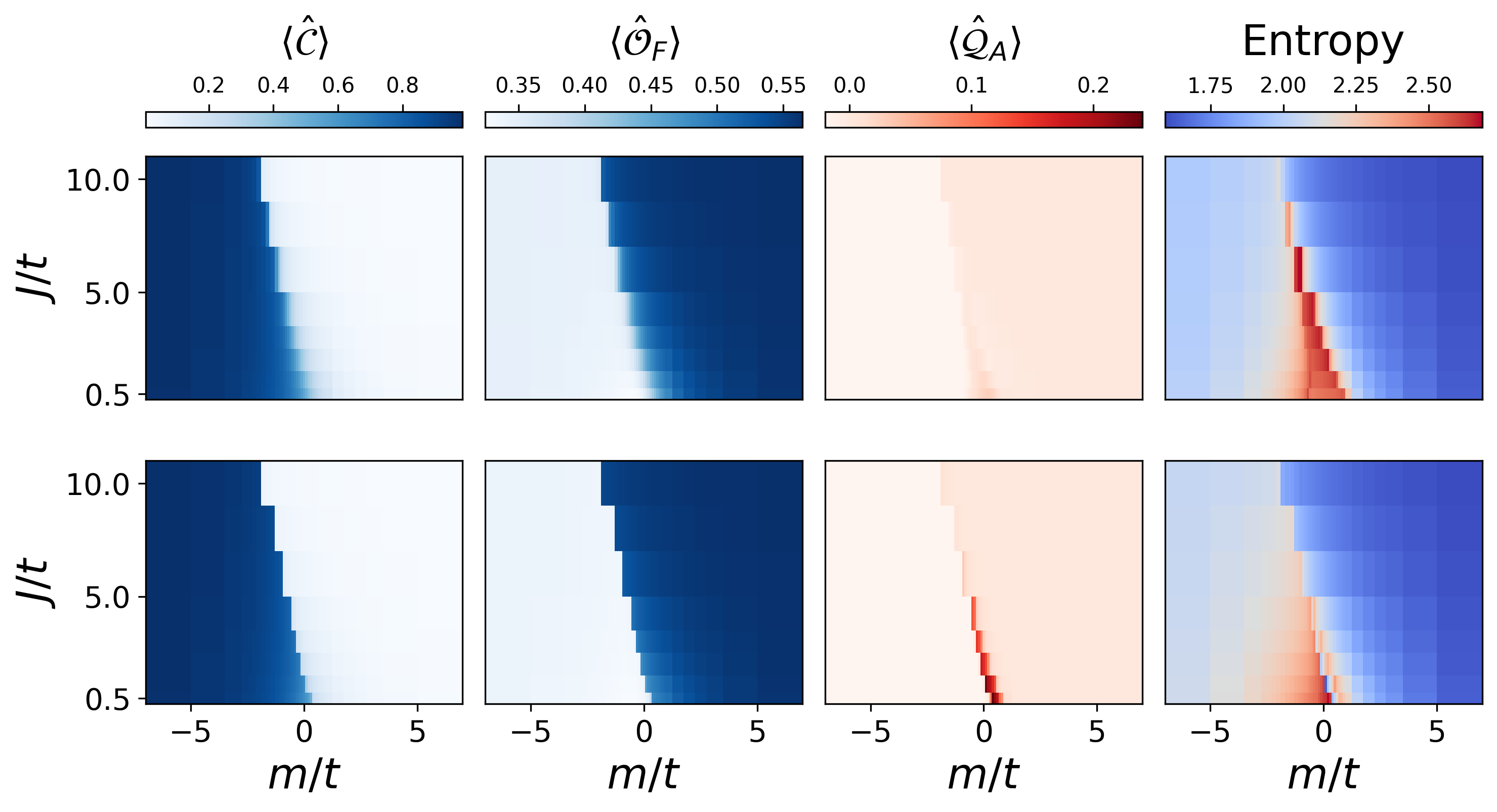}
    \caption{Physical observables for fermions (top panel) and bosons (bottom panel) are shown for $J/t>0$. Both fermions and bosons exhibit a phase transition from a dimer phase to a quantum link phase, with notable differences around $m/t=0$. The chiral condensate $\langle\hat{\mathcal{C}}\rangle$ and $\langle\hat{\mathcal{O}}_F\rangle$ are similar for both species, though the transition is sharper for bosons. In contrast, entropy and $\langle\hat{\mathcal{Q}}_A\rangle$ display more pronounced changes near the transition.}
    \label{phase}
\end{figure}

\textbf{\emph{Model.---}} 
We study the $2+1$D $\mathrm{U}(1)$ QLM on a square lattice with gauge fields coupled to matter fields described by the following Hamiltonian \cite{Hashizume2022},
\begin{align}\nonumber
\hat{H}=&-t\sum_{\mathbf{j},\mu=x,y}\Big(s_{\mathbf{j},\mathbf{e}_{\mu}}\hat{\psi}^{\dagger}_{\mathbf{j}}\hat{U}_{\mathbf{j},\mathbf{e}_{\mu}}\hat{\psi}_{\mathbf{j}+\mathbf{e}_{\mu}}+\textrm{h.c.}\Big)\\\label{eq:QLM}
&+m\sum_{\mathbf{j}}s_{\mathbf{j}}\hat{\psi}^{\dagger}_{\mathbf{j}}\hat{\psi}_{\mathbf{j}}-J\sum_{\Box}\Big(\hat{U}_{\Box}+\hat{U}^{\dagger}_{\Box}\Big).
\end{align}
The first term describes the minimal coupling of the matter fields $\hat{\psi}_\mathbf{j}$ (fermions or bosons) that live at the lattice sites $\mathbf{j}=(j_x,j_y)^\intercal$ to the gauge fields $\hat{U}_{\mathbf{j},\mathbf{e}_{\mu}}$ located at the links that connect the sites $\mathbf{j}$ and $\mathbf{j}+\mathbf{e}_{\mu}$, and the second term sets the mass of the matter fields. In this formulation, the hopping and mass terms are staggered by adopting the Kogut--Susskind framework. Staggering in the hopping term depends on the direction, where $s_{\mathbf{j},\mathbf{e}_{x}}=+1$ and $s_{\mathbf{j},\mathbf{e}_{y}}=(-1)^{j_{x}}$, which defines a translation symmetry up to two lattice spacings. Single lattice translations define the shift (chiral) symmetry \cite{Hashizume2022}.
The staggering in the mass term is given by $s_{\mathbf{j}}=(-1)^{j_x+j_y}$, which maps to a positive charge when the fermion/boson is present on an even site ($s_{\mathbf{j}} =1$) and a negative charge when the fermion/boson is present on an odd site ($s_{\mathbf{j}}= -1$). The last term of Eq.~\eqref{eq:QLM} is a plaquette contribution that describes magnetic energy, where $\hat{U}_{\Box}=\hat{U}_{\mathbf{j},\mathbf{e}_x}\hat{U}_{\mathbf{j}+\mathbf{e}_x,\mathbf{e}_y}\hat{U}^{\dagger}_{\mathbf{j}+\mathbf{e}_y,\mathbf{e}_x}\hat{U}^{\dagger}_{\mathbf{j},\mathbf{e}_y}$. The magnetic term is non-zero only for plaquettes that have a clear orientation and these are termed flippable plaquettes. We restrict the Hilbert space of the gauge fields by mapping them onto spin-$1/2$ operators with the choice $\hat{U}_{\mathbf{j},\mathbf{e}_{\mu}}=\hat{S}^+_{\mathbf{j},\mathbf{e}_{\mu}}$ for the gauge fields and $\hat{S}_{\mathbf{j},\mathbf{e}_{\mu}}^z$ for the local electric field. In this Letter, we are interested in the ground-state physics in the physical sector $\{\ket{\phi}\}$ with no background charges, which corresponds to the Gauss-law condition $\hat{G}_{\mathbf{j}}|\phi\rangle=0$, where $\hat{G}_{\mathbf{j}}$ is the Gauss's law operator given by,

\begin{align}\label{gauss}
\hat{G}_{\mathbf{j}}=\hat{\psi}^{\dagger}_\mathbf{j}\hat{\psi}_\mathbf{j}{-}\frac{1{-}({-}1)^\mathbf{j}}{2}{-}\sum_{\mu}\big(\hat{E}_{\mathbf{j},\mathbf{e}_{\mu}}{-}\hat{E}_{\mathbf{j}-\mathbf{e}_{\mu},\mathbf{e}_{\mu}}\big).
\end{align}

The model in Eq.~\eqref{eq:QLM} has been the subject of theoretical proposals for quantum simulation \cite{osborne2022largescale,surace2023abinitio}, and a version where the gauge fields are represented by spin-$1$ operators has recently been realized on a $3\times4$ square lattice on a \texttt{Quantinuum} quantum computer \cite{crippa2024analysisconfinementstring2}. Understanding in what regimes the matter statistics will lead to substantively different physics is crucial in guiding future experiments. The phase diagram, which we map out in the following, is a powerful tool in this endeavor.

\begin{figure}
\centering
    \includegraphics[width=\linewidth]{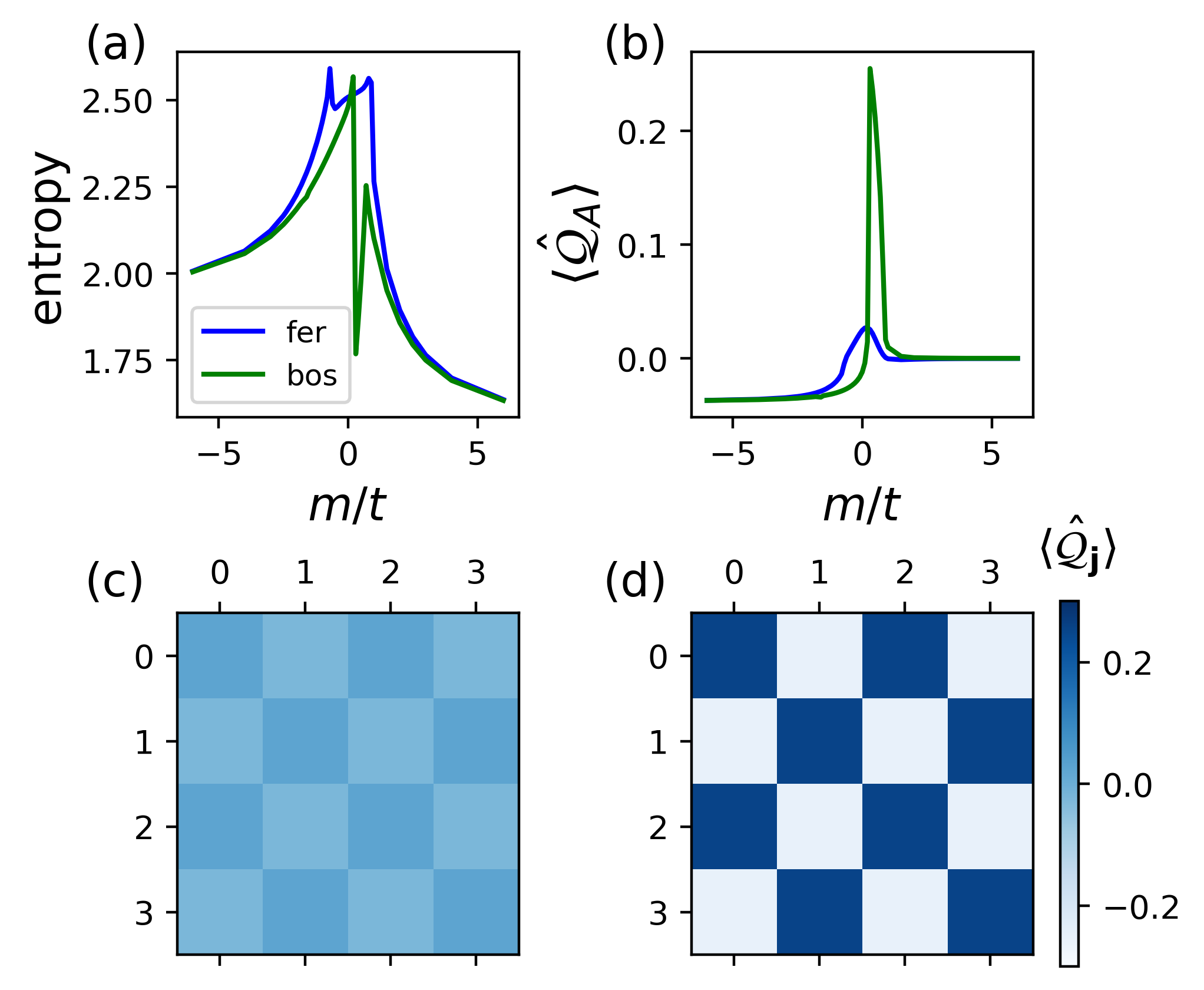}
    \caption{(a) The entanglement entropy and (b) $\langle\hat{\mathcal{Q}}_A\rangle$ are plotted for fermions and bosons across a range of $m/t$ values at $J/t=0.5$. Notably, the behavior near $m/t=0$ differs between bosons and fermions. The fact that the entanglement entropy drops discontinuously for bosons suggests a first order like phase transition in contrast to fermions. The profile of the order parameter $\langle\hat{\mathcal{Q}}_j\rangle$ is shown for (c) fermions and (d) bosons at $m/t=0.3$, $J/t=0.5$. It is observed that bosons tend to favor ordering more than fermions, where the plaquettes alternate in orientation. This behavior corresponds to a peak in $\langle\hat{\mathcal{Q}}_A\rangle$ and a dip in the entanglement entropy.}
    \label{ferbos}
\end{figure}

\textbf{\emph{Phase diagram.---}}
Since the physical system involves gauge and matter fields, one requires order parameters that are sensitive to the corresponding degrees of freedom. The system has a chiral symmetry and the order parameter that is sensitive to this symmetry is the chiral condensate,
\begin{align}\label{eq:CC}
\hat{\mathcal{C}}=\frac{1}{L_xL_y}\sum_{\mathbf{j}}(-1)^\mathbf{j}\bigg[\hat{n}_j-\frac{1-(-1)^\mathbf{j}}{2}\bigg],
\end{align}
where we consider an $L_x\times L_y$ lattice, with $L_x$ sites in the $x$-direction and $L_y$ sites in the $y$-direction. The chiral condensate~\eqref{eq:CC} corresponds to the matter degrees of freedom and is sensitive to charge--anti-charge fluctuations. To probe the ordering of gauge fields, we additionally study the local operators that are sensitive to the flippability and orientation of the plaquettes, which are given by 

\begin{subequations}
\begin{align}
&\hat{\mathcal{O}}_{\mathbf{j}}=\sum_{\mathbf{\eta=\pm}}\hat{P}^{\eta}_{\mathbf{j},\mathbf{e}_x}\hat{P}^{\eta}_{\mathbf{j}+\mathbf{e}_x,\mathbf{e}_y}\hat{P}^{\bar{\eta}}_{\mathbf{j}+\mathbf{e}_y,\mathbf{e}_x}\hat{P}^{\bar{\eta}}_{\mathbf{j},\mathbf{e}_y},\\
&\hat{\mathcal{Q}}_{\mathbf{j}}=\sum_{\mathbf{\eta=\pm}}\eta\hat{P}^{\eta}_{\mathbf{j},\mathbf{e}_x}\hat{P}^{\eta}_{\mathbf{j}+\mathbf{e}_x,\mathbf{e}_y}\hat{P}^{\bar{\eta}}_{\mathbf{j}+\mathbf{e}_y,\mathbf{e}_x}\hat{P}^{\bar{\eta}}_{\mathbf{j},\mathbf{e}_y},
\end{align}
\end{subequations}
where $\hat{P}^{\pm}_{\mathbf{j},e_x}=\frac{1}{2}\pm \hat{S}^{z}_{\mathbf{j},e_x}$ and $\bar{\eta}$ denotes a sign opposite to $\eta$. These operators acting on a plaquette are non-zero only when the corresponding plaquette is flippable. Note that while $\hat{\mathcal{O}}_{\mathbf{j}}$ is sensitive only to the flippability of a plaquette, $\hat{\mathcal{Q}}_{\mathbf{j}}$ is also sensitive to the orientation of the plaquette giving either positive or negative signs for clockwise and anticlockwise orientations, respectively. In order to probe ordering in the gauge fields, we can now define the following order parameters

\begin{align}
&\hat{\mathcal{O}}_F=\frac{1}{L_xL_y}\sum_{\mathbf{j}}\hat{\mathcal{O}}_{\mathbf{j}},\;\;\;\hat{\mathcal{Q}}_A=\frac{1}{L_xL_y}\sum_{\mathbf{j}}(-1)^{\mathbf{j}}\hat{\mathcal{Q}}_{\mathbf{j}}.
\end{align}
For perfect antiferromagnetic plaquette ordering, $\hat{\mathcal{Q}}_A$ is maximized; while for the RVB phase, $\hat{\mathcal{O}}_F$ has a finite contribution but $\hat{\mathcal{Q}}_A$ vanishes in the thermodynamic limit. In the columnar phase, $\hat{\mathcal{O}}_F \leq 0.5$, but  $\hat{\mathcal{Q}}_A \approx 0$. Thus, these order parameters allow the distinction of different phases that arise in the model \cite{Hashizume2022}.

We numerically compute ground states through iDMRG \cite{white1992,white1993,mcculloch2008,Uli_review} using the \texttt{Matrix Product Toolkit} \cite{mptoolkit} on cylinders of circumference $L_y=4$, with infinite axial length ($L_x\to\infty$). For spin models, this system size is known to approach the thermodynamic limit in two spatial dimensions in and out of equilibrium \cite{Hashizume2022,Hashizume2022DPTs,Hashizume2020Hybrid}. We enforce Gauss's law condition by adding a large energy penalty proportional to $\sum_{\mathbf{j}}\hat{G}_\mathbf{j}^2$ to our Hamiltonian. To ensure convergence, we increase the bond dimensions to as much as $\chi=800$ \cite{SM1}. In Fig.~\ref{phase}, we plot the order parameters for different choices of $m/t$ and $J/t$ for both the fermionic and bosonic matter fields. In the limit of $m/t\to -\infty$, the ground state exhibits broken chiral symmetry reflected in the saturation of the chiral condensate order parameter $\langle\hat{\mathcal{C}}\rangle$ to a non-zero value. As $m/t$ increases towards $+\infty$, the order parameter smoothly approaches $0$, with a sharp drop around $m/t=0$. The drop is more pronounced for bosons than for fermions. The gauge fields exhibit more complex ordering. Near $m/t=0$, where the matter hopping is enhanced, they tend to disorder the gauge fields but eventually lead to an ordered phase, which is detected by the order parameter $\langle \hat{\mathcal{O}}_F\rangle$ as $m/t$ is tuned to positive values. Notably, this ordering can occur even in the absence of, or with a weak, $J/t$ term, where the role of $J/t$ is primarily to further favor the ordering \cite{Hashizume2022}.

\begin{figure*}
\centering
\includegraphics[width=\linewidth]{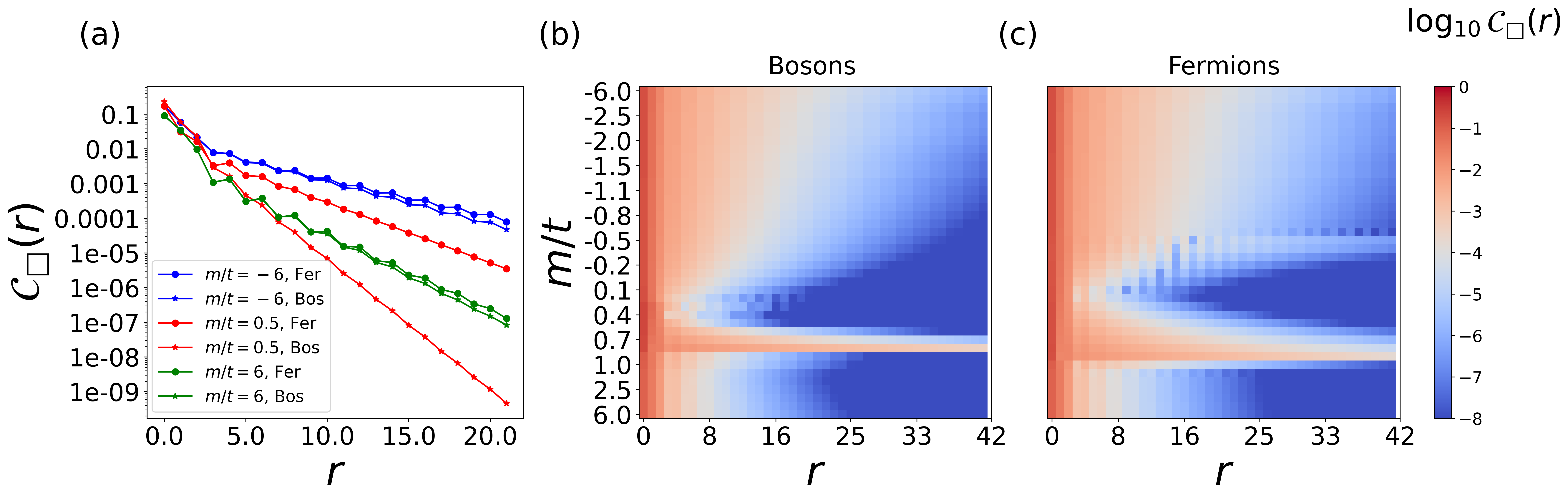}
\caption{(a) The behavior of connected plaquette correlation function Eq.~\eqref{plqc} as a function of distance $r$ from $\mathbf{j}=(0,0)$ for fermionic and bosonic cases for the choice $J/t=0.5$ and few $m/t$ values. Deep in the columnar and RVB phases, correlations behave similarly and are closely aligned for both bosons and fermions while they behave distinctly close to the transition. We do not present data for values very close to $m/t=0$, as the correlations exhibit rapid fluctuations that obscure the underlying behavior.  (b,c) The correlation function for the full range of $m/t$ values for bosons and fermions. Around the transition, correlations extend over longer ranges, with this region being narrower in the bosonic case than in the fermionic case, potentially indicating a spin-liquid phase.}
\label{plqcor}
\end{figure*}

\textbf{\emph{Bosons vs.~fermions.---}} When the mass term $\hat{H}_m$ and the plaquette term $\hat{H}_J$ approach zero in strength, the kinetic energy of the matter field is significant, entering a regime where the statistics of fermions and bosons are expected to play an important role. As shown in Fig.~\ref{phase}, around the region of small magnetic interactions $J/t$ and near $m/t=0$, we observe distinct deviations in the order parameters and entropy between fermions and bosons. The order parameters $\langle\hat{\mathcal{C}}\rangle$ and $\langle\hat{\mathcal{O}}_F\rangle$ exhibit comparable behavior for fermions and bosons, although the transition is notably sharper in the bosonic case. However, significant differences arise in the behavior of the entropy and the order parameter $\langle\hat{\mathcal{Q}}_A\rangle$ around the transition region. Comparable differences are also evident along a fixed $m/t$ cross section in the vicinity of the transition \cite{SM1}. To better understand this, we present the behavior of entropy and $\langle\hat{\mathcal{Q}}_A\rangle$ as functions of $m/t$ in Fig.~\ref{ferbos}(a,b) for $J/t=0.5$. Notably, for bosons, we observe a sharp drop in entropy near $m/t=0.3$ indicative of a potential first-order transition, although the possibility of a second-order transition cannot be definitively excluded \cite{SM1}. This is accompanied by a pronounced peak in the gauge field order parameter $\langle\hat{\mathcal{Q}}_A\rangle$, indicating that bosons tend to favor significant ordering, wherein the plaquettes alternate in orientation, compared to fermions and we confirm this by plotting the profile of $\langle\hat{\mathcal{Q}_\mathbf{j}}\rangle$ in Fig.~\ref{ferbos}(c,d).

To gain deeper insight into the nature of ordering in the gauge fields, we study the behavior of the connected plaquette correlation function

\begin{align}\label{plqc}
\mathcal{C}_{\Box}(r)=\langle \hat{P}^+_{\Box,\,\mathbf{j}} \hat{P}^{p(r)}_{\Box,\,\mathbf{j}+r {\mathbf{e}_x}}\rangle - \langle \hat{P}^+_{\Box,\,\mathbf{j}} \rangle \langle \hat{P}^{p(r)}_{\Box,\,\mathbf{j}+r {\mathbf{e}_x}}\rangle,
\end{align}
where $r$ is the separation from the plaquette at $\mathbf{j}=(0,0)$ and $\hat{P}^\pm_{\Box,\mathbf{j}}=\hat{P}^\pm_{\mathbf{j},\mathbf{e}_x}\hat{P}^\pm_{\mathbf{j}+\mathbf{e}_x,\mathbf{e}_y}\hat{P}^\mp_{\mathbf{j}+\mathbf{e}_y,\mathbf{e}_x}\hat{P}^\mp_{\mathbf{j},\mathbf{e}_y}$. When $r$ is even (odd), its parity is $p(r)=\pm$. As is evident from Fig.~\ref{plqcor}(a), deep within both the columnar and RVB phases, the gauge field correlations for fermions and bosons exhibit similar behavior. However, upon approaching the regime where the matter hopping is enhanced ($m/t\approx 0$), distinct differences arise. This region highlights the crucial role of particle statistics.
More importantly, the region of $m/t\gtrsim1$ where confinement physics is expected, both fermionic and bosonic correlations remain short-ranged and closely aligned. Interestingly, near the transition between these phases, correlations become significantly long-ranged, hinting at the emergence of a spin-liquid-like phase which appears somewhat narrower for bosons than for fermions as shown in Fig.~\ref{plqcor}(b,c). This is also consistent with the understanding of Ref.~\cite{Ram2022}, which demonstrated that the bosonic vs.~fermionic nature of the degrees of freedom is not important until the exchange statistics become important.

\textbf{\emph{Simulation and experiments.---}} 
Bosons offer several advantages over fermions in the simulation and experimental realization of QLMs. One key advantage is that Monte Carlo simulations involving fermions are hindered by the notorious sign problem \cite{Troyer2005computational}, whereas bosonic simulations circumvent this issue, enabling more efficient numerical approaches. Additionally, hardcore bosons can be digitally simulated on quantum computers through a straightforward mapping onto qubits, enhancing their accessibility for quantum computational methods. Furthermore, bosons are naturally suited for cold-atom experiments, making the physical implementation of QLMs more feasible \cite{aidelsburger2021cold,Halimeh_review}. Recent proposals suggest realizing QLMs by mapping spinless bosons onto an optical superlattice while enforcing gauge symmetry through a linear gauge protection term \cite{Halimeh2020e,osborne2023spins}, which hold promise for extensions to $2+1$D \cite{osborne2022largescale}. In contrast, simulating QLMs with fermionic matter necessitates a significantly more complex quantum algorithm involving a sequence of quantum gate operations on qubits. This approach requires the application of the Jordan--Wigner transformation, which introduces non-locality and presents substantial challenges in maintaining computational efficiency. However, it is worth mentioning that recent methods such as compact encoding show a promising alternative route to this problem \cite{jafarizadeh2024recipelocalsimulationstronglycorrelated,Derby2021compact,nigmatullin2024experimentaldemonstrationbreakevencompact}. When it comes to cold-atom realizations, fermionic matter requires spin-dependent potentials with significant engineering overhead in order to locally enforce Gauss's law \cite{surace2023abinitio}.

\textbf{\emph{Conclusions.---}}  
In this work, we explored the ground-state phase diagram of the $\mathrm{U}(1)$ quantum link model with hardcore-bosonic matter and performed a comparative analysis with its fermionic counterpart. Our results show that the bosonic phase diagram closely mirrors that of fermions, exhibiting a transition from a columnar phase to an RVB phase. However, notable differences arise in the vicinity of the phase transition, where matter degrees of freedom have enhanced mobility, and particle statistics significantly influence the phase. In particular, near the transition, bosons, unlike fermions, tend to favor gauge-field ordering within a narrow region, possibly followed by a thin liquid-like regime before entering the confined phase, whose physical origin should be examined in detail. In the confined phases, both bosonic and fermionic cases display remarkably similar behavior in terms of correlation functions and order parameter profiles. 

Interestingly, even in the regime of small $J/t$ and $m/t\approx0$, where matter statistics plays a nontrivial role, our findings show that the choice of observable is also important. Although the chiral condensate and nonstaggered gauge-field order parameters do not reveal any qualitative difference between bosons and fermions, the staggered gauge-field order parameter and the entanglement entropy uncover a more nuanced picture indicating that matter statistics plays an important role.

These findings not only prompt deeper theoretical and numerical exploration into the role of quantum statistics in lattice gauge theories, but also have practical implications. From an experimental standpoint, especially in simulating confinement physics, our results suggest that bosons can effectively replicate fermionic behavior, making them a viable alternative in both analog and digital quantum simulation platforms. In particular, it shows that the implementation of fermionic matter degrees of freedom in string breaking experiments in $2+1$D lattice gauge theories may not be crucial. These experiments tend to focus on the confined regime \cite{cochran2024visualizingdynamicschargesstrings,gonzalezcuadra2024observationstringbreaking2,crippa2024analysisconfinementstring2}. As shown in this work, this is where hardcore-bosonic and fermionic matter degrees of freedom bring about the same physics.

\medskip

{\textbf{\textit{Acknowledgments.---}}The authors are grateful to Tomohiro Hashizume and Philipp Hauke for discussions in the early stages of this project. N.S.S., J.J.O., and J.C.H.~acknowledge funding by the Max Planck Society, the Deutsche Forschungsgemeinschaft (DFG, German Research Foundation) under Germany’s Excellence Strategy – EXC-2111 – 390814868, and the European Research Council (ERC) under the European Union’s Horizon Europe research and innovation program (Grant Agreement No.~101165667)—ERC Starting Grant QuSiGauge. This work is part of the Quantum Computing for High-Energy Physics (QC4HEP) working group.}

\appendix

%


\clearpage
\widetext
\begin{center}
\textbf{\large Supplemental Material for `Bosonic vs.~Fermionic Matter in Quantum Simulations of $2+1$D Gauge Theories'}
\end{center}
\makeatletter
\renewcommand{\c@secnumdepth}{0}
\makeatother
\setcounter{equation}{0}
\setcounter{figure}{0}
\setcounter{table}{0}
\setcounter{page}{1}
\makeatletter
\renewcommand{\theequation}{S\arabic{equation}}
\renewcommand{\thefigure}{S\arabic{figure}}
\renewcommand{\thesection}{S\arabic{section}}
\section{Physical observables and entropy along the $m/t=0$ cross-section}

\begin{figure}[H]
\centering
\includegraphics[width=10cm]{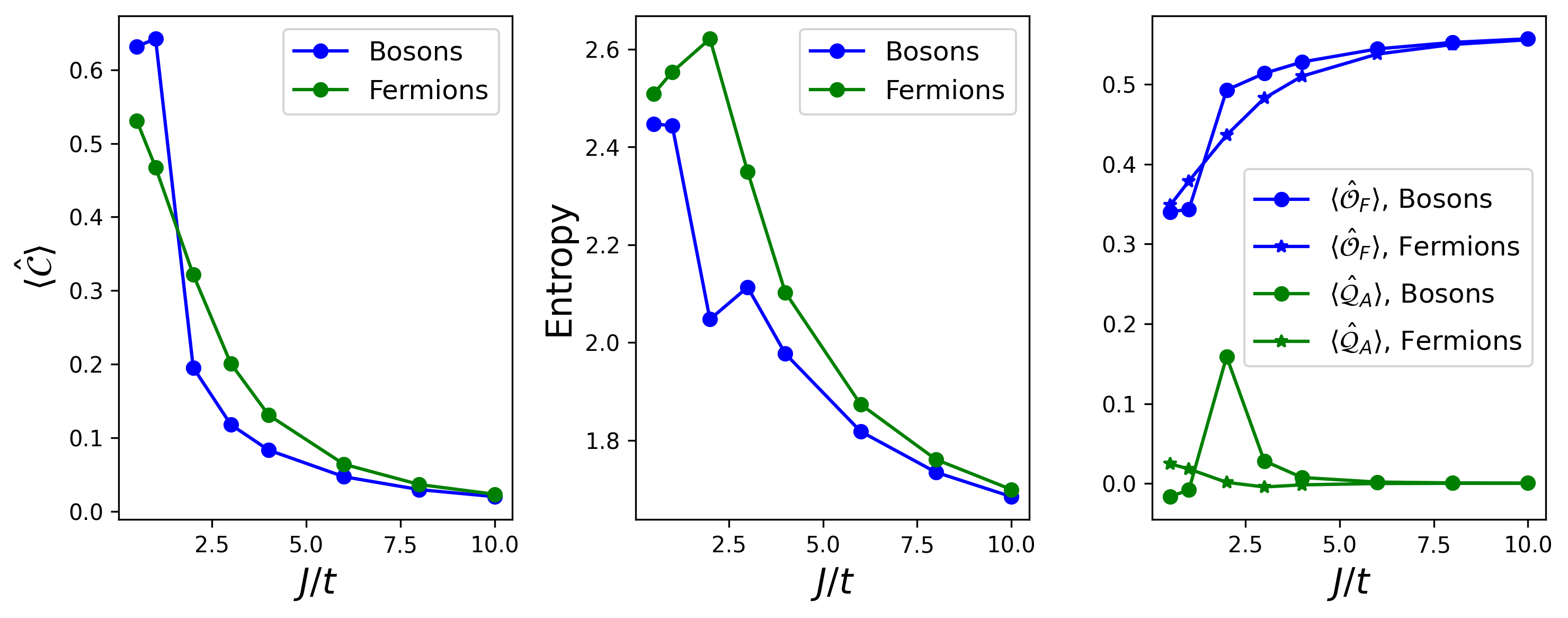}
\caption{Physical observables and entropies plotted for the QLM with fermionic and bosonic matter along the $m/t=0$ cross section.}
\label{acrossj}
\end{figure}

Here we present the order parameters and entropies for both bosons and fermions along the $m/t=0$ cross-section, as shown in Fig.~\ref{acrossj}. For large values of $J/t$, the behavior of bosons and fermions is similar, with data points closely aligned, which is expected since we are in the regime of strong magnetic interactions where the matter fields become effectively frozen. However, distinct differences emerge in the regime of small $J/t$. Notably, the entropy for bosons exhibits a dip around $J/t=2$ which is reflected in the sharp transition in the order parameters $\langle\hat{\mathcal{C}}\rangle$, $\langle\hat{\mathcal{O}}_F\rangle$ and $\langle\hat{\mathcal{Q}}_A\rangle$.

\section{Entanglement scaling for bosons}

\begin{figure}[H]
\centering
    \includegraphics[width=8cm]{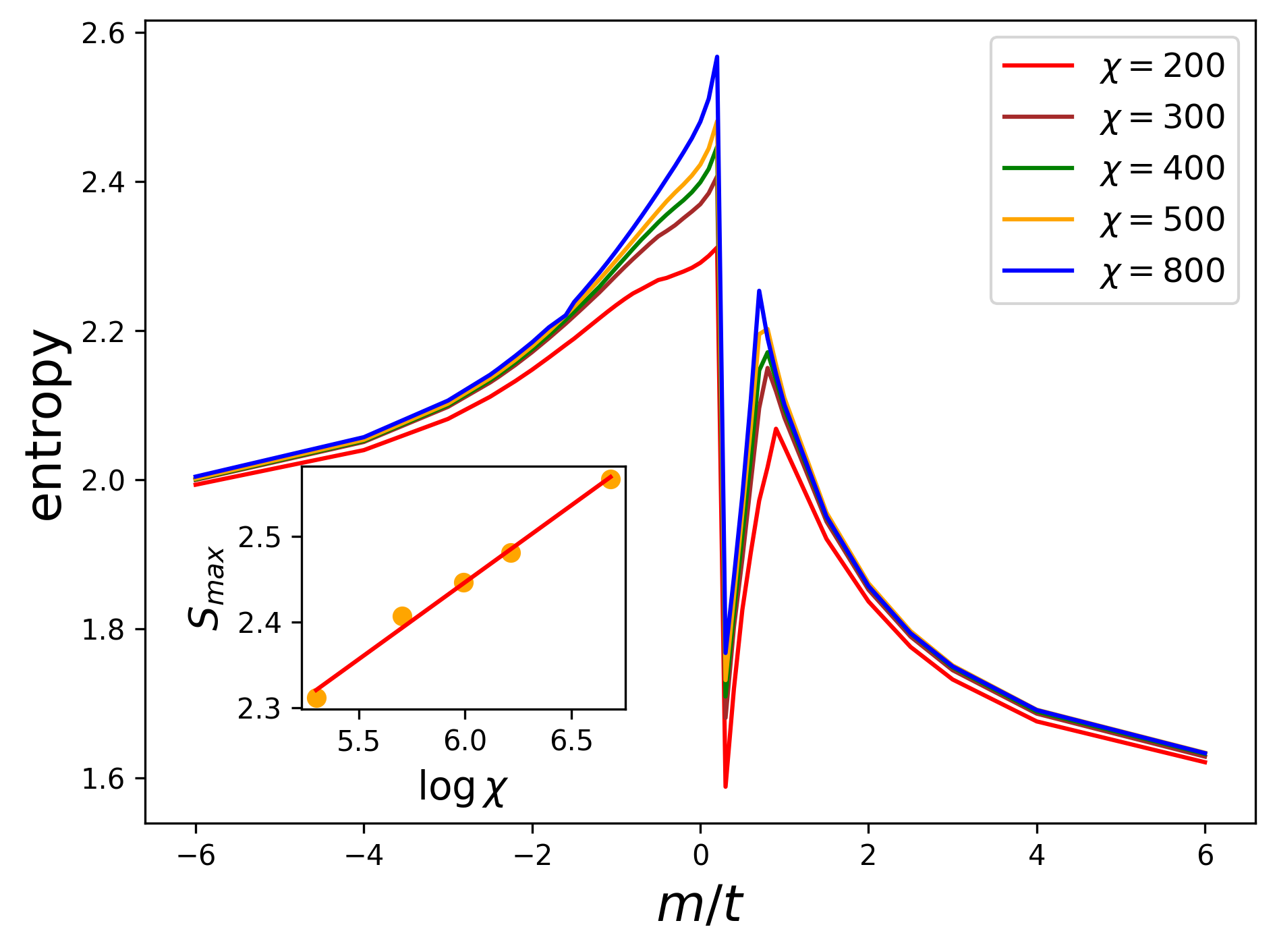}
    \caption{Entanglement entropies plotted against $m/t$ for a range of bond dimensions $\chi$ for the QLM with bosonic matter at $J/t=0.5$. Inset shows the scaling of the peak value of entanglement entropy $S_{\mathrm{max}}$ with $\log \chi$ at $m/t=0.2$.}
    \label{ent}
\end{figure}

Phase transitions in finite-sized systems are typically characterized by analyzing the scaling behavior of order parameters with system size. In the context of infinite systems, this role is effectively played by the finite bond dimension $\chi$ used to approximate the ground state via a matrix product state for a given set of Hamiltonian parameters \cite{Tagliacozzo}. Notably, near the critical point, the entanglement entropy exhibits a logarithmic scaling with the bond dimension, given by

\begin{equation} S_{\chi} = c \log{\chi}, \end{equation}
where $c$ is a constant that captures the universal scaling behavior. In Fig.~\ref{ent} we plot the entanglement entropy as a function of $m/t$ for a range of bond dimensions $\chi$ at $J/t=0.5$. We observe that while the entanglement saturates with $\chi$ deep in the columnar and RVB phases, it shows an increasing trend close to the phase transition. To probe its precise scaling, we plot its peak value at $m/t=0.2$ against $\chi$ and we find that it shows the scaling behavior $S_{\mathrm{max}}\approx \frac{1}{6}\log \chi$ for the range of $\chi$ we investigate hinting towards a possibility of a second-order transition \cite{Nagy_2011}. However, this warrants a further detailed study to understand the precise nature of transition.

\section{$L_y=2$ case}

\begin{figure}[H]
\centering
    \includegraphics[width=8cm]{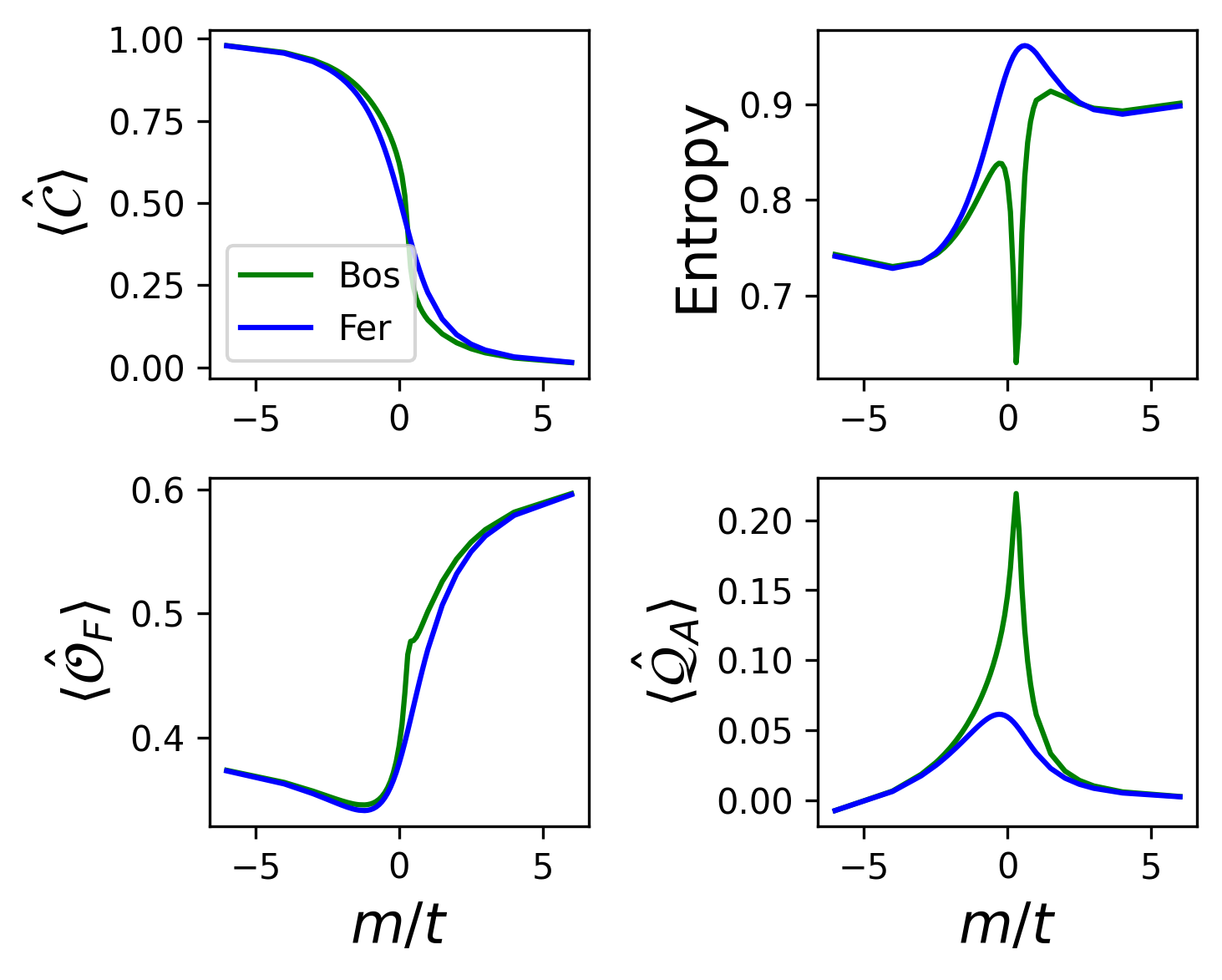}
    \caption{Physical observables and entropy plotted against $m/t$ for fermions and bosons at $J/t=0.5$ for a cylinder of diameter $L_y=2$.}
    \label{L2}
\end{figure}

In one spatial dimension, fermions and hardcore bosons are indistinguishable due to the absence of particle exchange freedom, leading to an expectation of similar phase diagrams for the quantum link model (QLM) in 1D. However, this equivalence breaks down as one moves away from strictly one-dimensional systems, where particle statistics begin to play a significant role. To investigate this, we numerically examine the differences in observable behavior and entanglement entropy for fermions and bosons on a cylindrical geometry with transverse width $L_y=2$. As shown in Fig.~\ref{L2}, notable differences emerge near $m/t=0$ between the two cases, particularly in the entanglement entropy and the gauge field order parameter $\langle\hat{\mathcal{Q}}_A\rangle$. Similar to the $L_y=4$ case, the bosonic system exhibits a pronounced dip in entropy and a corresponding peak in $\langle\hat{\mathcal{Q}}_A\rangle$ near $m/t=0$. In contrast, the fermionic system displays a smoother variation in entropy and $\langle\hat{\mathcal{Q}}_A\rangle$ near $m/t=0$.

\end{document}